\documentclass{article}
\usepackage{amsmath}
\usepackage{amssymb}
\usepackage{bm}
\usepackage{booktabs}
\usepackage{graphicx}
\usepackage{multirow}
\usepackage[figuresleft]{rotating}
\usepackage{url}
\begin{document}
\title{PETE: A Hybrid Photonic and Electronic Thermoelectric Device for Direct Heat-to-Electricity Conversion}
\author{Yasmine Ben Mechlia$^{1}$, Shuo Chen$^{2}$, and Bo Zhao$^{1,*}$\\[0.5em]
$^{1}$Department of Mechanical and Aerospace Engineering\\
University of Houston, Houston, TX 77204\\
$^{2}$Department of Physics and Texas Center for Superconductivity\\
University of Houston, Houston, TX 77204\\
$^{*}$Corresponding author: \url{bzhao8@uh.edu}}
\date{}
\maketitle

\begin{abstract}
Photonic heat engines can achieve high conversion efficiencies, but their far-field power density is constrained by the blackbody limit. Near-field devices overcome this limitation through photon tunneling across a nanoscale vacuum gap; however, the mechanical spacers required to maintain this gap introduce parasitic heat conduction. We propose PETE (Photonic and Electronic Thermoelectric), a hybrid solid-state architecture in which thermoelectric generator (TEG) legs replace passive spacers. The legs simultaneously maintain the near-field gap and convert conducted heat into electricity. We evaluate PETE using an indium tin oxide emitter, an indium arsenide photovoltaic cell, and TaFeSb- and ZrNiSn-based Half-Heusler thermoelectric materials. At a hot-side temperature of 977~K, the device achieves a power density of 18.6~W$\cdot$cm$^{-2}$ at 19.76\% efficiency. This power density is approximately twice that of either the standalone near-field thermophotovoltaic device or the standalone TEG, while the efficiency is 44\% higher than that of the standalone TEG. The performance can be tuned through thermoelectric geometry and material selection, and its adaptability is demonstrated for a broad range of thermoelectric materials and operating temperatures. PETE provides a general strategy for recovering spacer-mediated heat leakage in near-field photonic heat engines.
\end{abstract}

\section{\label{sec:introduction}Introduction}

Waste heat is one of the most abundant yet underutilized energy resources in modern industrial systems. The need to harvest it has become increasingly urgent with the recent boom of gigawatt-scale electricity consumption from AI data centers and similar emerging loads, driving the demand for higher-performance waste heat recovery technologies. Photonic heat engines represent a class of solid-state heat-to-electricity converters that require no moving parts and are, in principle, scalable across a wide range of temperatures and heat source configurations. Their major advantage is their efficiency, sometimes exceeding 40\%, compared to other solid-state solutions such as thermoelectrics \cite{cui_nonreciprocal_2026,lapotin_thermophotovoltaic_2022,zhao_chemical_2020,papadakis_broadening_2020,zhao_high-performance_2017}. However, in the far-field heat transfer regime, where the distance between the heat source and heat sink is in the macroscale, the power density produced by photonic heat engines is constrained by the blackbody limit. To overcome this, zero-vacuum gap designs \cite{habibi_enhanced_2025}, thermo-photo-thermionic \cite{rahman_micro-gap_2023} devices and thermophotonic systems \cite{zhao_self-sustaining_2019,harder_thermophotonics_2003} were proposed, each achieving at least twice the power density of the regular photonic heat engines. However, these achievements are still considered low compared to near-field photonic heat engines, or are comparable but require extremely high-quality materials \cite{zhao_near-field_2018}.

\begin{figure}
\centering
\includegraphics[width=1\columnwidth,page=1]{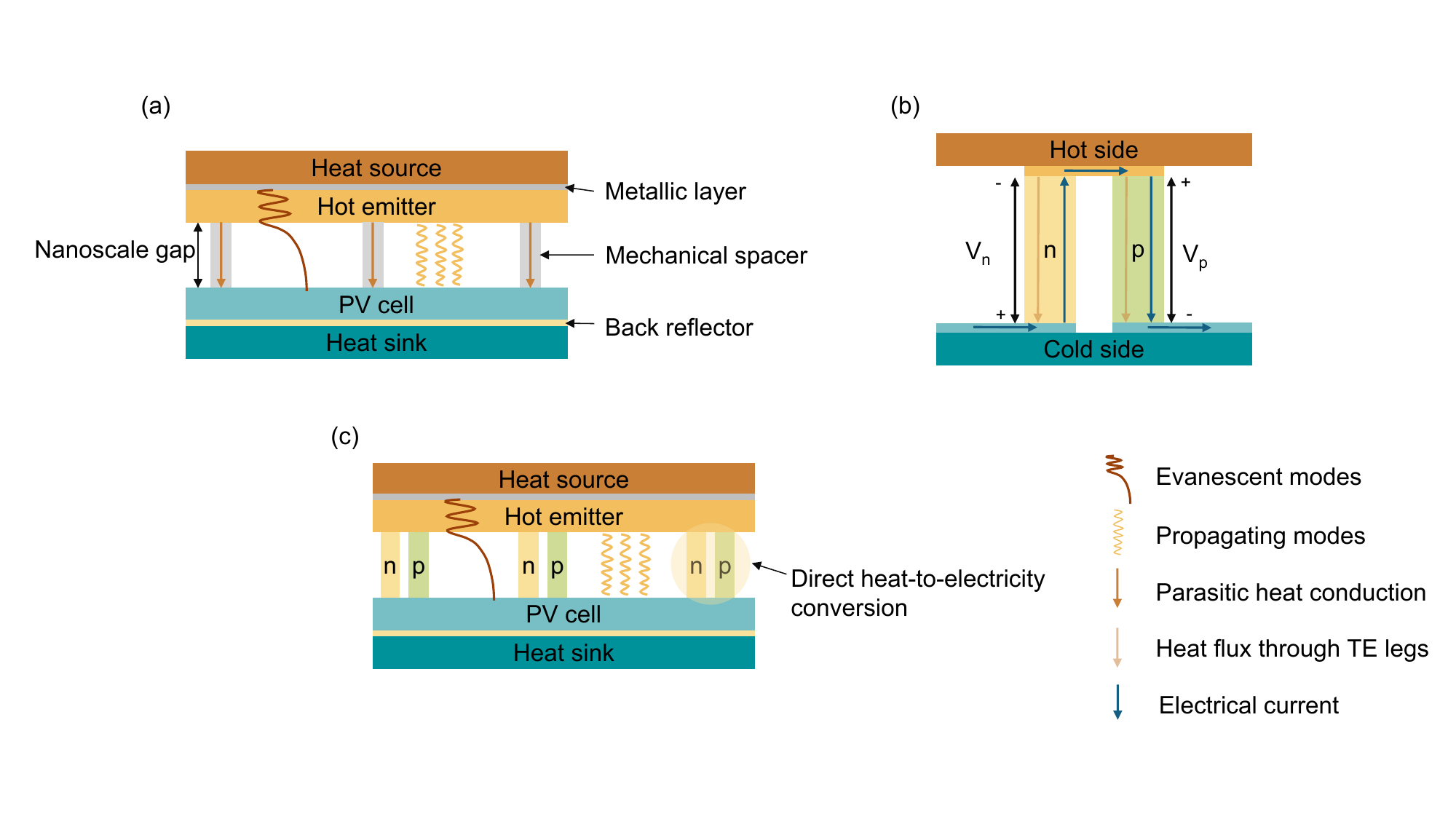}
\caption{Hybridization principle of the PETE device. (a) Architecture of traditional NFTPV cells with mechanical spacers. (b) Schematic of a basic thermoelectric generator, which generates a voltage, V$_p$-V$_n$, under a temperature difference. (c) Architecture of the PETE device, where TEGs maintain the nanoscale gap required by the NFTPV, converting the temperature difference directly into electrical power.}
\label{fig:spacers}
\end{figure}

Near-field photonic heat engines exceed the blackbody limit using geometry: reducing the gap distance between the heat source and the heat sink to the nanoscale. In this way, they access the evanescent modes, only observable in the subwavelength scales and achieve power densities orders of magnitude higher than the far-field regime.
However, maintaining sub-wavelength gap distances across macroscopic device areas is challenging, and demonstrated approaches using piezoelectric actuation \cite{mittapally_near-field_2021} or active NEMS/MEMS systems \cite{bhatt_integrated_2020} could only be applicable to small areas. Passive mechanical spacers have emerged as the most practical and widely adopted solution \cite{loubet_micronanoscale_2025} because they can be fabricated using available technologies, and provide structural rigidity and the ability to maintain uniform sub-micron gaps across chip-scale surfaces. However, due to the direct contact, the spacers conduct heat directly from the hot emitter to the cold receiver without contributing to electrical power generation (Fig.~\ref{fig:spacers}(a)), significantly constraining the efficiency of near-field photonic heat engines \cite{selvidge_large_2025, shi_optimized_2021, tang_near-field_2020, desutter_near-field_2019, inoue_one-chip_2019, watjen_near-field_2016, hu_near-field_2008, dimatteo_micron-gap_2004,dimatteo_enhanced_2001}.

We propose to resolve this limitation by replacing passive spacer materials with thermoelectric generators (TEGs), which convert the temperature difference at their junctions directly into electricity (Figs.~\ref{fig:spacers}(b) and \ref{fig:spacers}(c)). We use near-field thermophotovoltaics (NFTPV) to demonstrate our idea, and unlike prior proposals \cite{shan_parametric_2021,chubb_combined_2018}, our technology integrates the thermoelectric element directly inside the vacuum gap. The resulting hybrid device --- called PETE (Photonic and Electronic Thermoelectric) --- harvests near-field radiative heat and converts the conducted heat flux through the thermoelectric legs. We demonstrate that by optimizing material geometry, PETE can outperform both NFTPV and TEG, yielding record-breaking performance.

In the first section of this paper, we develop the theoretical framework governing PETE performance, progressing from an ideal constant-property model to a realistic temperature-dependent model. In the second section, we numerically evaluate the performance of this technology on an
ITO--InAs NFTPV system and study the influence of thermoelectric geometry and temperature dependence. The impact of thermal and electrical contact resistances is analyzed, and adaptability is demonstrated across a broad catalog of thermoelectric materials and operating temperatures. The final section addresses model limitations and identifies directions for future work.

\section{\label{sec:modeling}System Modeling}

\begin{figure}
\centering
\includegraphics[width=0.80\textwidth,page=1]{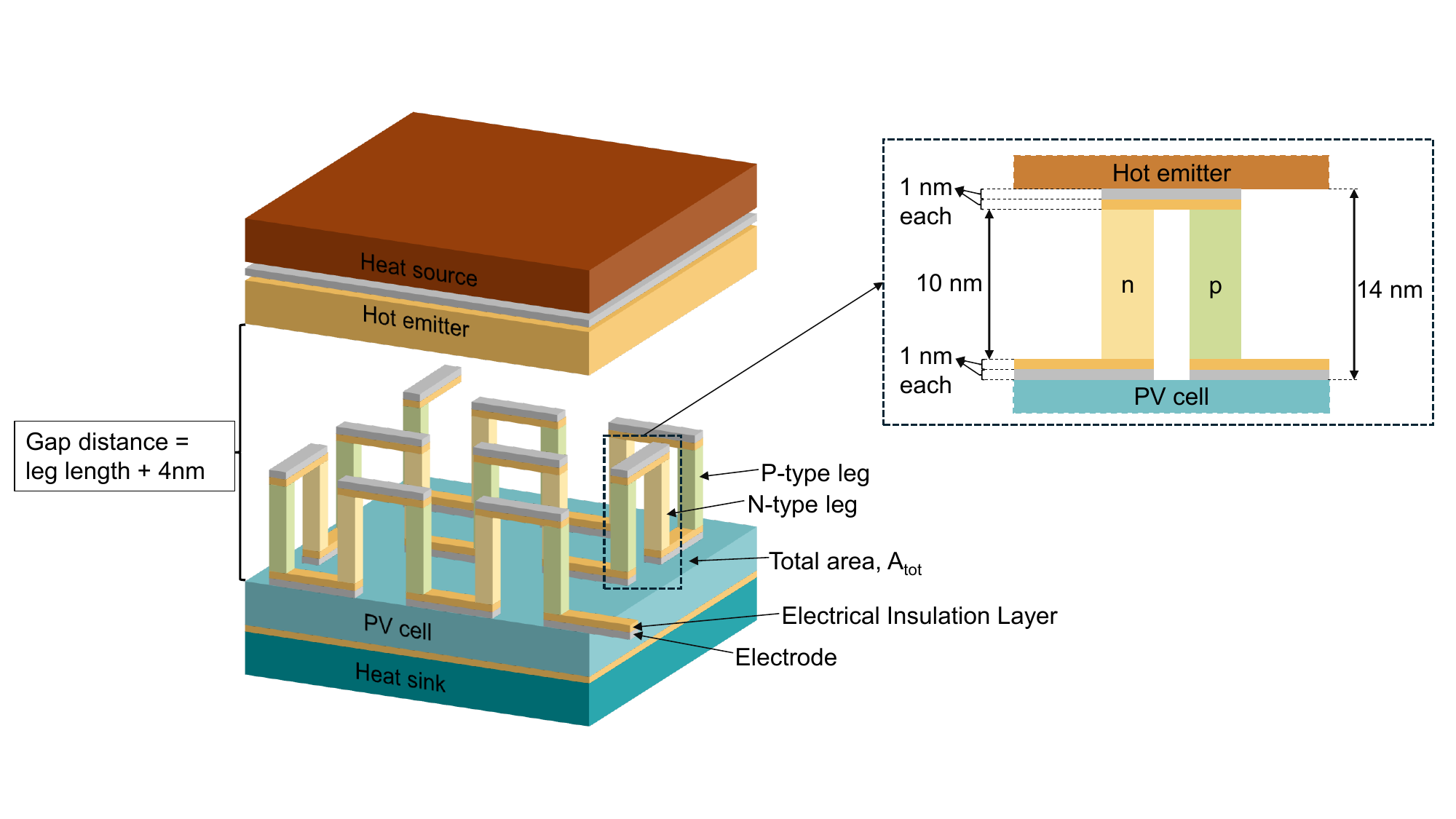}
\caption{3D structure of the PETE device: $n$- and $p$-type thermoelectric legs, connected by electrodes and separated by electrical insulation layers, are arranged in a periodic array across the surface of the emitter and the PV cell. The gap distance equals the leg length plus 4~nm, accounting for the electrode and insulation layer thicknesses (2~nm) at each junction.}
\label{fig:3D}
\end{figure}

The main goal of this section is to benchmark the performance for the PETE technology, which is the joint contribution of the TEG spacer system and the photonic heat engine (Eqs.~\ref{eq:pete-power-density}--\ref{eq:pete-efficiency}). We therefore proceed by evaluating the performance of each subsystem in isolation.
Here, we assume diffuse thermal and electronic transport, so that the traditional thermal and electrical resistance circuits can be used. A general single-junction NFTPV cell architecture consisting of a thin-film thermal emitter and photovoltaic cell is considered. The emitter and PV cell are separated by a vacuum gap in the range of 10~nm to 1~$\mu$m. The TEG system (Fig.~\ref{fig:spacers}b) is composed of thermoelectric legs with identical, uniform, and constant cross-sections, electrodes for electric current conduction, and electrical insulation layers to prevent current leakage (Fig.~\ref{fig:3D}). The TEG spacer system performance is detailed below. The performance of the NFTPV is summarized in the Results section.

\subsection{\label{sec:ideal-model}Ideal Performance Model of PETE}

We first focus on obtaining the ideal performance of the PETE device, which requires an optimal performance of the TEG spacer system.
We assume ideal electrodes and insulation layers (no electrical or thermal contact resistance) to allow exposure to the full temperature difference. The performance of the TEG system is then estimated using the constant property model, where the thermoelectric properties are taken at the mean temperature \cite{dunham_thermoelectric_2019,dunham_power_2015,baranowski_response_2014}, and are considered to be symmetric for the $n$ and $p$ type legs, i.e. identical electrical and thermal conductivities and Seebeck coefficients of equal magnitude but opposite signs.
The energy balance at the hot and cold junctions accounts
for the Peltier, Joule and conduction heat fluxes (Eqs.~\ref{eq:hot-side-heat-flux-cpm}--\ref{eq:peltier-hot-heat-flux-cpm}). Joule heating is a volumetric effect generated by the material itself, split identically and supplied to both junctions of both TE legs as an approximation. This implies that it should be subtracted from the total heat flux, thus the $\frac{1}{2}$ coefficient (Eq.~\ref{eq:joule-heat-flux-cpm}).

Electrically, since the TEG legs are connected in series (and thermally in parallel), the system is modeled as a real generator with internal resistance equal to that of the legs, feeding power to an external load \cite{dunham_power_2015,min_new_2022}(Fig.~\ref{fig:figure3_combined}).
The electrical performance of the TEG can be optimized by varying the load resistance, $R_L$, or the ratio $m = R_L/R$, called the load-matching factor.
Depending on the value of $m$, the TEG can either work in maximum power mode ($m=1$) or maximum efficiency mode ($m = \sqrt{1+ZT}$), where $ZT$ is the thermoelectric figure of merit, purely defined from material properties \cite{freer_key_2022}. This yields the set of Eqs.~\ref{eq:pete-power-density}--\ref{eq:peltier-hot-heat-flux-cpm}:

\begin{figure}
\centering
\includegraphics[width=0.80\textwidth,page=1]{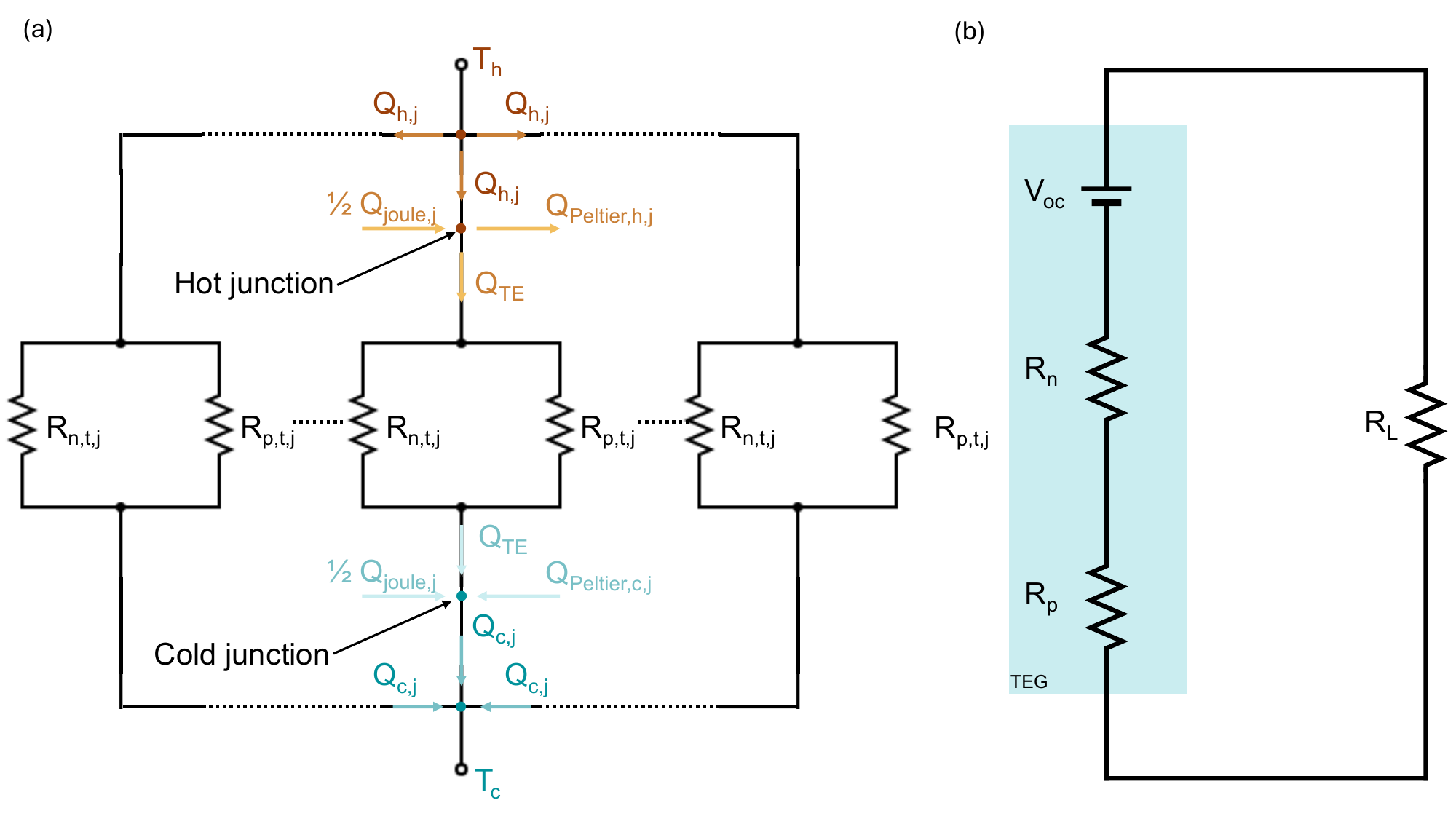}
\caption{Schematic of (a) Thermal resistance model of the TEG spacers with the input, output, Joule, Peltier and conduction heat fluxes, $Q_{h,j}$, $Q_{c,j}$, $Q_{Joule,j}$, $Q_{Peltier,h,j}$, $Q_{Peltier,c,j}$ and $Q_{TE,j}$, where $R_{n,p,t,j}$ represent the thermal resistances of each $n$- or $p$-type leg. $T_h$ and $T_c$ are the temperatures of the hot and cold junctions, and of the emitter and PV cell simultaneously. (b) Electrical Model of the TEG spacer system where $R_n$ and $R_p$ represent the total electrical resistances of the $n$-type and $p$-type legs, respectively.}
\label{fig:figure3_combined}
\end{figure}

\begin{equation}
 p_{PETE} = p_{TEG} + p_{NFTPV}
\label{eq:pete-power-density}
\end{equation}

\begin{equation}
\begin{aligned}
\eta_{PETE} &= \frac{p_{TEG}+p_{NFTPV}}{q_{rad}+q_h} \\
&= \eta_{TEG}\frac{q_{h}}{q_{rad}+q_h}
+ \eta_{NFTPV}\frac{q_{rad}}{q_{rad}+q_h}.
\end{aligned}
\label{eq:pete-efficiency}
\end{equation}

where:
\begin{equation}
p_{TEG} = \frac{2nA}{A_{tot}} \frac{S^2 \Delta T^2}{d\, \rho} \frac{m}{(m + 1)^2}
\label{eq:teg-power-density-cpm}
\end{equation}
\begin{equation}
q_h = q_{peltier,h} + q_{TE} - \frac{1}{2}q_{Joule}
\label{eq:hot-side-heat-flux-cpm}
\end{equation}
\begin{equation}
\eta_{TEG} = \frac{p_{TEG}}{q_h}
\label{eq:teg-efficiency-cpm}
\end{equation}
\begin{equation}
q_{TE}=2\,n\,A\,\frac{\kappa}{d\,A_{tot}}\,\Delta T
\label{eq:conduction-heat-flux-cpm}
\end{equation}
\begin{equation}
\frac{1}{2}q_{Joule}=\frac{nA}{A_{tot}}\;\frac{S^{2}\Delta T^{2}}{(m+1)^{2}\,d\,\rho}
\label{eq:joule-heat-flux-cpm}
\end{equation}
\begin{equation}
q_{peltier,h}=\frac{2nA}{A_{tot}}\;\frac{S^{2}T_{h}\Delta T}{(m+1)\,d\,\rho}
\label{eq:peltier-hot-heat-flux-cpm}
\end{equation}
where:
\begin{equation}
A_n = A_p = A
\label{eq:equal-leg-areas}
\end{equation}

In Eqs.~\ref{eq:pete-power-density}--\ref{eq:equal-leg-areas}, $p$ is power density and $\eta$ is conversion efficiency, with subscripts PETE, TEG, and NFTPV identifying the subsystem (full device / thermoelectric subsystem / standalone near-field thermophotovoltaic subsystem). $q$ is a heat flux: subscript $h$ denotes the hot junction of the TEG, $rad$ the NFTPV radiative channel, $Peltier$ the Peltier contribution, $TE$ the thermoelectric-conduction contribution, and $Joule$ the Joule-heating contribution. The TEG geometry is described by $A$, the cross-sectional area of one leg ($A_n$, $A_p$ for the $n$-type and $p$-type legs, equal here by the ideal-model symmetry assumption); $d$, the leg length; $n$, the number of leg pairs; and $A_{tot}$, the total emitter area. The leg material is characterized by $S$, $\rho$, $\kappa$, the Seebeck coefficient, electrical resistivity, and thermal conductivity. $\Delta T = T_{h}-T_{c}$ is the temperature drop, with subscripts $h$ and $c$ for the hot (emitter-side) and cold (PV-side) junctions.

The length scale of the TE legs implies that the heat conduction is orders of magnitude larger than the thermal radiation from the emitter. Therefore, the total area of the TE legs must remain very small compared to the overall NFTPV surface area; thus, the radiation heat flux can be treated as a constant in the calculations. However, the performance of the TEG also depends on the cross-sectional area of the legs, creating a trade-off with heat conduction suppression. Therefore, a geometrical parameter known as the fill fraction, or $FF$, is defined to control this tradeoff (Eq.~\ref{eq:fill-fraction}):
\begin{equation}
FF = \frac{n(A_p + A_n)}{A_{tot}} = \frac{2nA}{A_{tot}}
\label{eq:fill-fraction}
\end{equation}

The fill fraction $FF$ encapsulates both the number of leg pairs $n$ and the individual leg cross-sectional area $A$. For a given $FF$, $n$ and $A$ are not independently constrained by the energy balance — only their product matters for performance. In practice, once $FF$ is fixed by the performance optimization, $n$ and $A$ can be chosen independently based on fabrication constraints and mechanical requirements \cite{dunham_thermoelectric_2019}.
To satisfy the tradeoff, the fill fraction must take low values, in the range of $10^{-7}$ to $10^{-4}$.

\subsection{\label{sec:temperature-dependent-model}Realistic Considerations: Temperature-Dependent Thermoelectric Properties}

Under large temperature differences, the assumptions for the ideal performance model may break down. In this case, thermoelectric materials are temperature-dependent, so the definition of the $ZT$, as well as all of the equations of the system will have to account for this. We have therefore chosen the cumulative temperature property model, where the properties are now integrated across the temperature domain instead of taking the value at the mean temperature. This model has been proven to be reasonably accurate and computationally cheap \cite{kim_relationship_2015,tuley_zt_2017}. The $ZT$ is then given by (Eq.~\ref{eq:zt-eng}) \cite{zhu_understanding_2019,tuley_zt_2017, kim_relationship_2015}:

\begin{equation}
(ZT)_{eng} = Z_{eng}\Delta T
=\frac{\left(\int_{T_c}^{T_h} S(T)dT\right)^2}{\int_{T_c}^{T_h} \rho(T)dT \int_{T_c}^{T_h} \kappa(T)dT} \Delta T
\label{eq:zt-eng}
\end{equation}

\begin{equation}
(PF)_{eng} = \frac{\left(\int_{T_c}^{T_h} S(T)dT\right)^2}{\left(\int_{T_c}^{T_h} \rho(T)dT\right)}
\label{eq:engineering-power-factor}
\end{equation}

where PF is the engineering power factor, describing the material's ability to extract electrical power from a temperature difference.

\begin{figure}
\centering
\includegraphics[width=1\columnwidth,page=1]{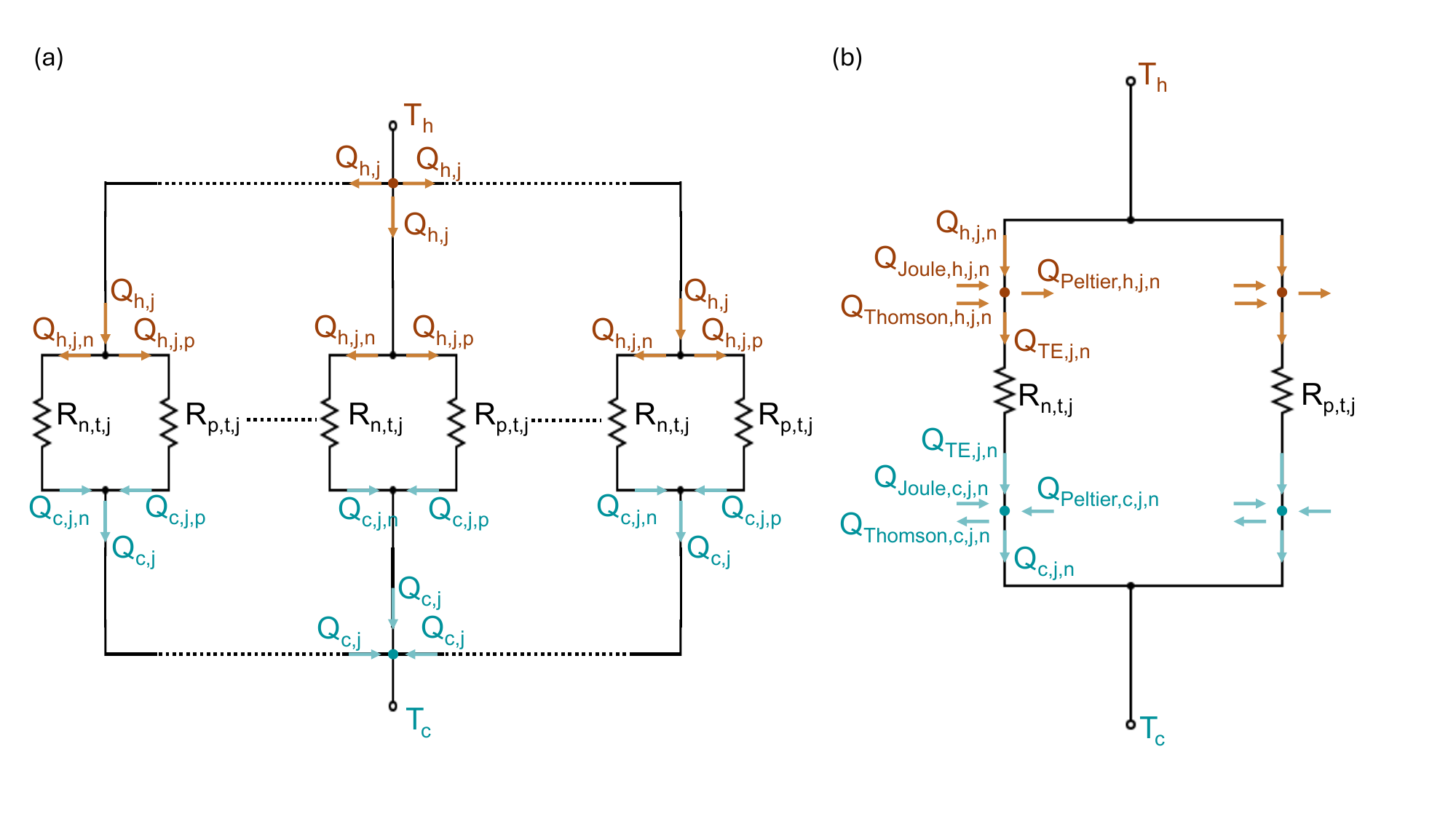}
\caption{(a) Thermal circuit showing heat flux distribution through the thermocouple array, where $Q_{h,j}$ and $Q_{c,j}$ are the total heat fluxes at the hot and cold junctions of each thermocouple, and $Q_{h,n,j}$, $Q_{h,p,j}$, $Q_{c,n,j}$, $Q_{c,p,j}$ are the individual contributions from the $n$- and $p$-type legs, respectively. $T_h$ and $T_c$ are the temperatures of the hot and cold junctions, and of the emitter and PV cell simultaneously. (b) Thermal circuit of a single thermoelectric couple showing the direction of the different heat fluxes at the hot and cold junctions.}
\label{fig:figure4_hot_cold_junction}
\end{figure}

The thermal resistance circuit of the device (Fig.~\ref{fig:figure4_hot_cold_junction}) shows the different heat fluxes present and their interactions at the hot and cold junctions. Thomson heating, also a volumetric effect, is now accounted for. Due to the asymmetric property distribution across TE legs, the weights assigned to the Thomson and Joule heating are now $W_T$ and $W_J$, respectively (See Supplementary Material, Eqs.~4--5 \cite{mechlia_supplementary_nodate}), instead of $\frac{1}{2}$. The electrical modeling is the same as in the ideal case. The general formula of the impedance matching coefficient as well as its expressions for the maximum power and efficiency configurations, and the general device $ZT_{d,eng}$ and efficiency are updated following the same derivations as in the ideal model, and in the literature in general (See Supplementary Material, Eqs.~1--18 \cite{mechlia_supplementary_nodate}).

To optimize the geometry of the TEG and to understand its impact on the performance of the PETE technology, all variables have been expressed as a function of the fill fraction $FF$ and the cross-sectional area ratio of the $n$- and $p$-type legs, denoted as $f$ (Eq.~\ref{eq:area-ratio}), accounting for the property asymmetry:
\begin{equation}
 f = \frac{A_n}{A_p}
\label{eq:area-ratio}
\end{equation}

To maximize the power density, we have verified that, in addition to $m=1$, $f$ should equal $\sqrt{\frac{\int_{T_c}^{Th}\rho_n\,dT}{\int_{T_c}^{Th}\rho_p\,dT}}$, which is consistent with the literature \cite{kim_efficiency_2015}. The maximum efficiency case is more complex to derive analytically so the $FF$ and $f$ will be numerically optimized.

\section{\label{sec:results}Results and Discussion}

To demonstrate our technology, we chose the ITO--InAs NFTPV system \cite{zhao_high-performance_2017}, with the state of the art performance for single-junction NFTPV: a power density of 11~W$\cdot$cm$^{-2}$ and an efficiency reaching 40\% for an emitter temperature of 900~K, a PV cell temperature of 300~K and a gap distance of 10~nm without spacers. For the purpose of our study, we have resimulated the NFTPV performance at a gap distance of 14~nm, taking into account the different components of the TEG spacer system. The thicknesses of the ITO emitter and the InAs cell are optimized to be 30~nm and 0.4~$\mu$m respectively.
The NFTPV is assumed to deliver the maximum power density, which is only slightly different from the maximum efficiency point \cite{zhao_high-performance_2017}.

\subsection{\label{sec:ideal-results}Ideal Performance Model}

To verify feasibility and understand the impact of geometry and material choice of the TE legs, we pick two examples of materials whose performance will be compared.

The first candidate material is the Yttrium-doped Mg$_3$Sb$_2$ semiconductor ($n$-type), with a peak $ZT$ of 1.16 at 759~K \cite{kihou_thermoelectric_2021}. The second candidate material is the oxide-purified, polycrystalline sodium-doped SnSe semiconductor ($p$-type), whose $ZT$ reaches a peak of 3.1 at 783~K \cite{zhou_polycrystalline_2021}. Their properties are reported in Table~\ref{tab:te_properties}.
The choice criterion is based on their $ZT$ as well as their thermal conductivity. To satisfy the symmetry assumption of the constant property model, each material is paired with a hypothetical mirror-image counterpart of the opposite carrier type, having identical electrical and thermal conductivity magnitudes and an opposite-sign Seebeck coefficient of equal magnitude.
These properties have been taken at 600~K, the mean between the temperatures of the emitter and the PV cell.

\begin{table}
\caption{\label{tab:te_properties}Thermoelectric properties of the candidate materials used in the ideal performance model.}
\centering
\resizebox{\columnwidth}{!}{%
\begin{tabular}{lcccc}
\toprule
\textrm{Material} & \textrm{$S_{600~K}$ ($\mu$V$\cdot$K$^{-1}$)} & \textrm{$\rho_{600~K}$ ($\Omega\cdot$m)} & \textrm{$\kappa_{600~K}$ (W$\cdot$m$^{-1}\cdot$K$^{-1}$)} & \textrm{$ZT_{\mathrm{600~K}}$} \\
\midrule
Mg$_{3.3}$Y$_{0.02}$Sb$_{1.5}$Bi$_{0.5}$ & -200 & $30\times10^{-6}$ & 1 & 0.925 \\
Na$_{0.03}$Sn$_{0.965}$Se & $\sim$ 325 & $1.215\times10^{-4}$ & $\sim$ 0.31 & 1.932 \\
\bottomrule
\end{tabular}
}
\end{table}

\begin{figure}[t]
\centering
\includegraphics[width=0.90\textwidth,page=1]{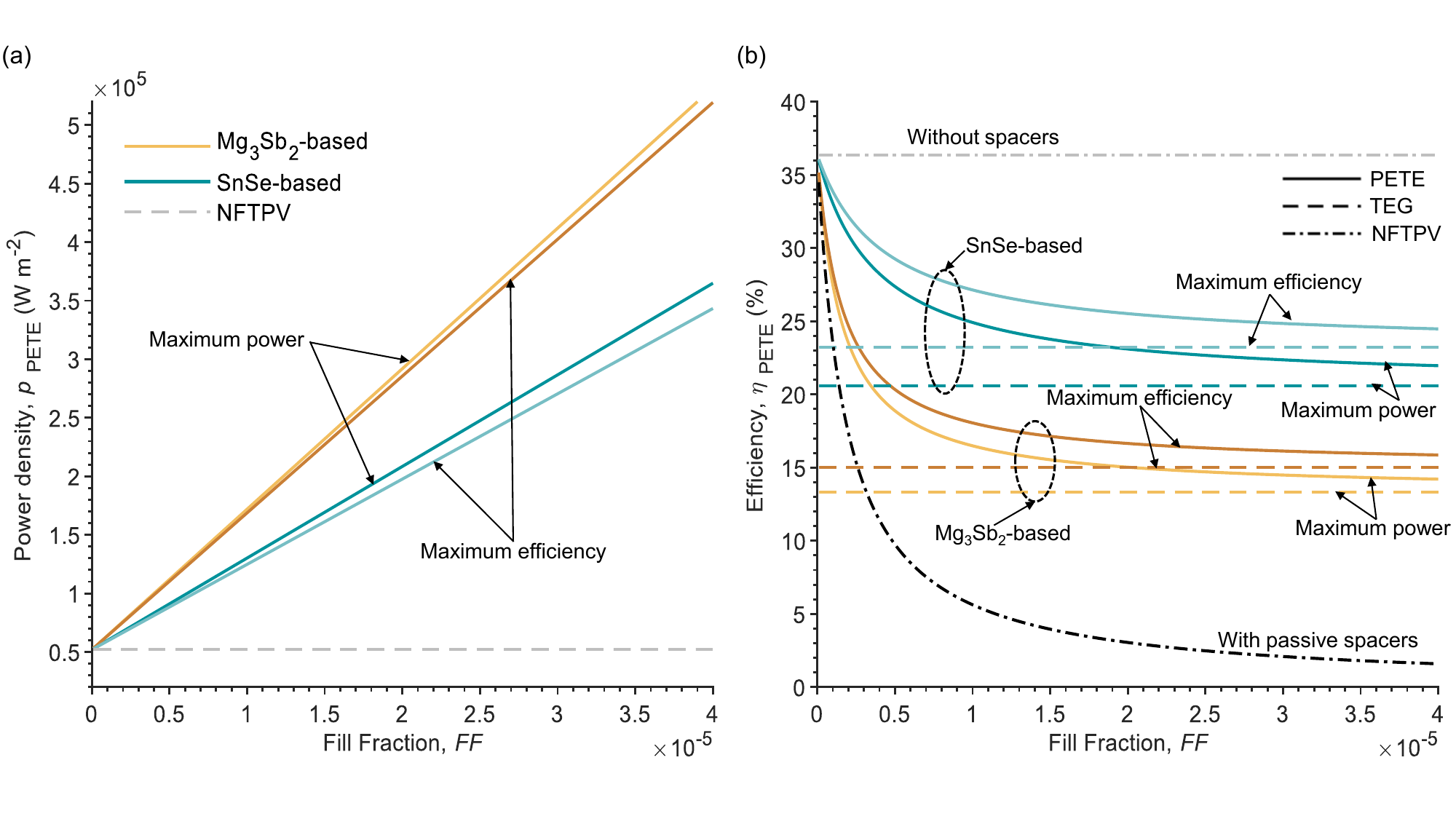}
\caption{Theoretical performance of the PETE device as a function of fill fraction at a leg length of 10~nm, for Mg$_3$Sb$_2$-based and SnSe-based thermoelectric materials. (a) Overall power density compared to the standalone NFTPV baseline. (b) Efficiency compared to the standalone NFTPV and TEG. As a reference, the NFTPV with passive $SiO_{2}$ spacers, with $\kappa_{SiO_{2}}=1.3~W/m \cdot {K}$ is represented by the black line.}
\label{fig:figure4_combined}
\end{figure}

Fig.~\ref{fig:figure4_combined} presents the resulting performance of the PETE device under an emitter temperature of 900~K.
It is obvious from Eq.~\ref{eq:pete-power-density} that the output power density is linearly proportional to the fill fraction $FF$.
However, what is more interesting is the relationship between $FF$ and the PETE efficiency.
$\eta_{PETE}$ decreases hyperbolically as a function of $FF$, and converges towards that of the TEG.
This shows that the thermoelectric materials should themselves be efficient to ensure the overall PETE efficiency remains high, according to Eq.~\ref{eq:pete-efficiency}.

It can be seen that the Mg$_3$Sb$_2$-based material produces a much higher power density than the SnSe-based material. This is due to the low electrical resistivity of the Mg$_3$Sb$_2$-based material which increases the electrical current and the power factor, PF. The SnSe-based material, however, has lower thermal conductivity and higher Seebeck coefficient than the former material, creating more balance between the different energy components within the TEG. Therefore, according to Eq.~\ref{eq:pete-efficiency}, increasing the TEG efficiency helps to flatten the curve of efficiency as a function of $FF$. This leads to higher $FF$ values and a larger geometry manipulation margin, and consequently, higher PETE performance values. This further proves the importance of choosing the right thermoelectric material for the right device conditions.

Thus, the criteria for choosing highly-performing thermoelectric materials for our application are: (1) high Seebeck coefficient, (2) low electrical resistivity, both of which can be summarized into a high power factor, $PF = \frac{S^2}{\rho}$, and (3) a low thermal conductivity. The ensemble of these criteria translates into a high figure of merit, which is a well-established criterion in literature. However, it is quite challenging to verify all three at once because the parameters themselves are interrelated via the charge carrier concentration \cite{shi_toward_2025}.

Another interesting consideration is the mechanical resistance of the TEG legs. Larger $FF$ margins mean the cross-sectional area could be adjusted such that the mechanical compression stresses are reduced, making the structure of the TEG more resistant \cite{merienne_thermal_2019}.
This becomes more important when comparing with the efficiency of the NFTPV with passive mechanical spacers. While all materials can reach high efficiencies at low fill fractions, the passive mechanical spacers case yields the lowest FF for the same efficiency, making the device fragile and unstable.

\subsection{\label{sec:realistic-results}Performance Model with Temperature Dependence and Property Asymmetry}

While the ideal performance model used Mg$_3$Sb$_2$- SnSe-based materials to set the material selection criteria, the realistic model adopts Half-Heusler materials. Although their figure of merit is lower than those of many others, they provide the thermomechanical durability required to sustain the nanoscale gap under large temperature differences \cite{shi_toward_2025,freer_key_2022}. In this second model, the material choice is particularly delicate because of the relative matching required between the $n$- and $p$-type legs. Therefore, the matching strategy begins by comparing the $ZT$ of candidate materials of the same type: their peak values and their variation with the temperature.
Preferably, the material should have a high peak value at high temperatures while increasing relatively steadily with temperature, which can be concluded from literature \cite{kim_relationship_2015}.
Once the material corresponding to these conditions is chosen, the next step is to approximately match it with another material of the other type, i.e. the order of magnitude of the properties should be comparable and the variation with respect to temperature should be similar.

For this model, a pair of Half-Heusler materials has been carefully chosen. The $n$-type is the ZrNiSn-based Half-Heusler for its high thermoelectric properties \cite{ai_high-performance_2025}.
The matching $p$-type is the TaFeSb-based Half-Heusler \cite{naderloo_performance_2025} (See Supplementary Material, Fig.~S2 \cite{mechlia_supplementary_nodate}).
The data provided for the $n$ and $p$ material do not share the same temperature points, therefore, an interpolation has been made to increase the number of integration points but also to match the maximum temperatures. As a result, the emitter and hot junction temperature are set at 977~K. Therefore, the performance of the NFTPV is re-evaluated at this temperature for the different TE leg lengths, all while accounting for the extra $4~nm$ contributed by the electrodes and electrical insulation layers.

Since the geometrical parameters are interrelated, and following the conclusions from the first model, the geometry optimization process aims to find $f$ at which the TEG reaches its maximum efficiency. Then depending on the user's end goal, $FF$ is determined accordingly.

\begin{figure}[t]
\centering
\includegraphics[width=0.90\textwidth,page=1]{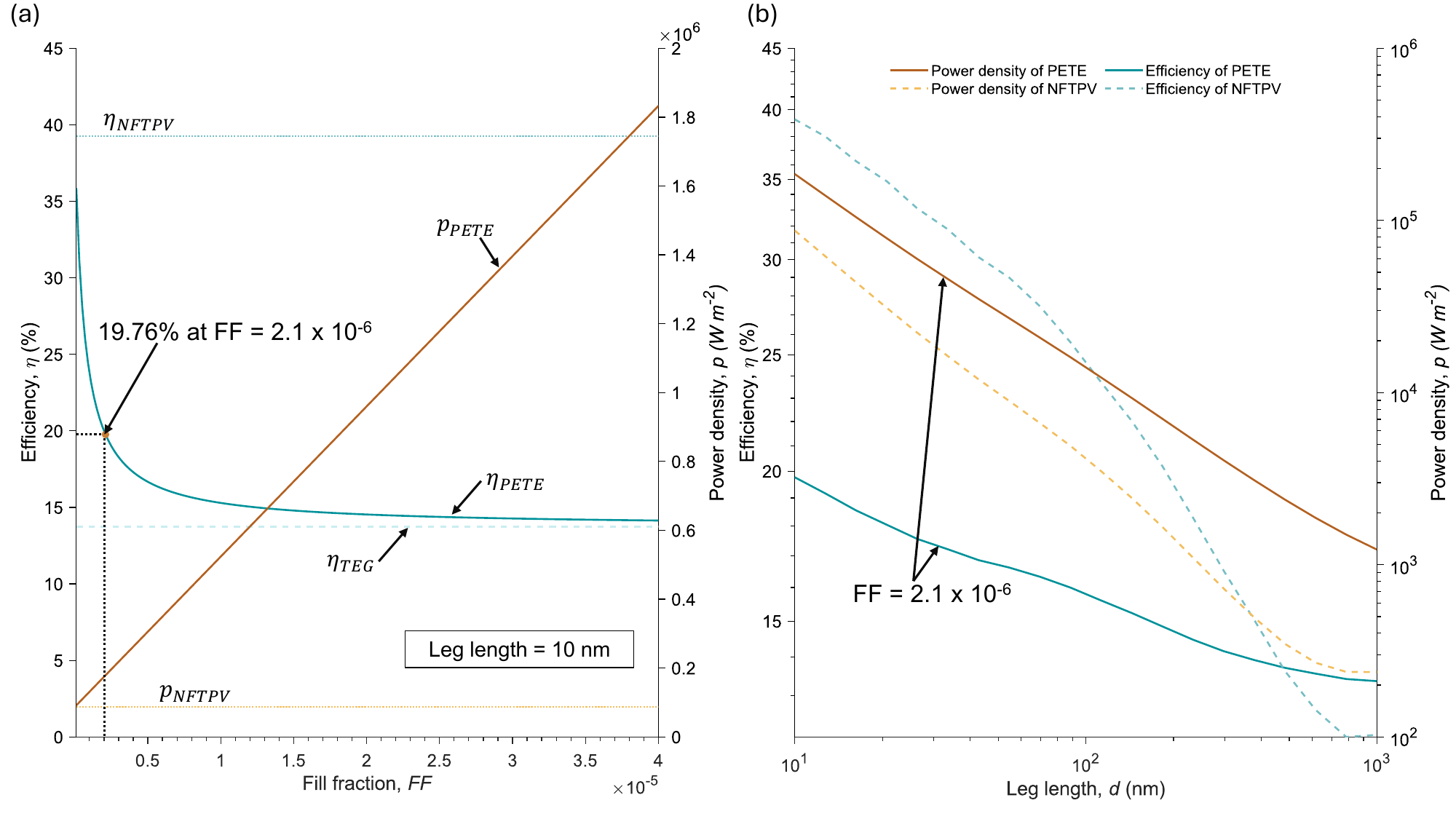}
\caption{Realistic performance of the PETE device compared to the standalone NFTPV, using TaFeSb- and ZrNiSn-based Half-Heusler thermoelectric materials at an emitter temperature of 977~K. (a) Power density and efficiency as a function of fill fraction $FF$, where the chosen operating point yields 19.76\% efficiency at $FF = 2.1\times 10^{-6}$. (b) Power density and efficiency as a function of leg length, compared to the standalone NFTPV, at the chosen $FF = 2.1 \times 10^{-6}$.}
\label{fig:figure5_leglength}
\end{figure}

Fig.~\ref{fig:figure5_leglength} clearly depicts the power density difference between the PETE and the NFTPV as a function of the leg length, demonstrating the superiority of the PETE device performance in comparison with the NFTPV. This proves that our technology still delivers very high power densities even as we increase the gap distance. We also observe the same overall trends of the power density and efficiency.

According to Eq.~\ref{eq:pete-efficiency}, aiming for around half the efficiency of the NFTPV lets us achieve a workable balance between $\eta$ and $p$ without collapsing the TEG contribution. At a $14~nm$ gap distance, the NFTPV converts heat to electricity with 39.26\% efficiency. At 19.76\% which is higher than recent TEG efficiencies achieved \cite{ghosh_high-entropy-driven_2024}, our technology reaches a power density of 18.6~W$\cdot$cm$^{-2}$, almost twice the amount of electrical power density the NFTPV generates by itself, for $FF = 2.1\times10^{-6}$ and $f=0.95$. To put this into perspective, the TEG produces 9.86~W$\cdot$cm$^{-2}$ at an efficiency of 13.73\%, while the NFTPV produces 8.73~W$\cdot$cm$^{-2}$, clearly demonstrating the immediate benefit and the tunability of the PETE technology.
If we closely examine Fig.\ref{fig:figure5_leglength}(b), we can see two other important features of the technology. First, the PETE efficiency is more stabilized as a function of the leg length, in comparison with the NFTPV, which guarantees a delivered power at a relatively high efficiency. Second, starting from $d=500~nm$, the efficiency of the PETE actually dominates the NFTPV. This means that PETE is still efficient even as we approach the far-field regime.

\subsection{\label{sec:adaptability}Adaptability to Temperature}

\begin{figure}[htbp]
\centering
\includegraphics[width=0.95\textwidth,page=1]{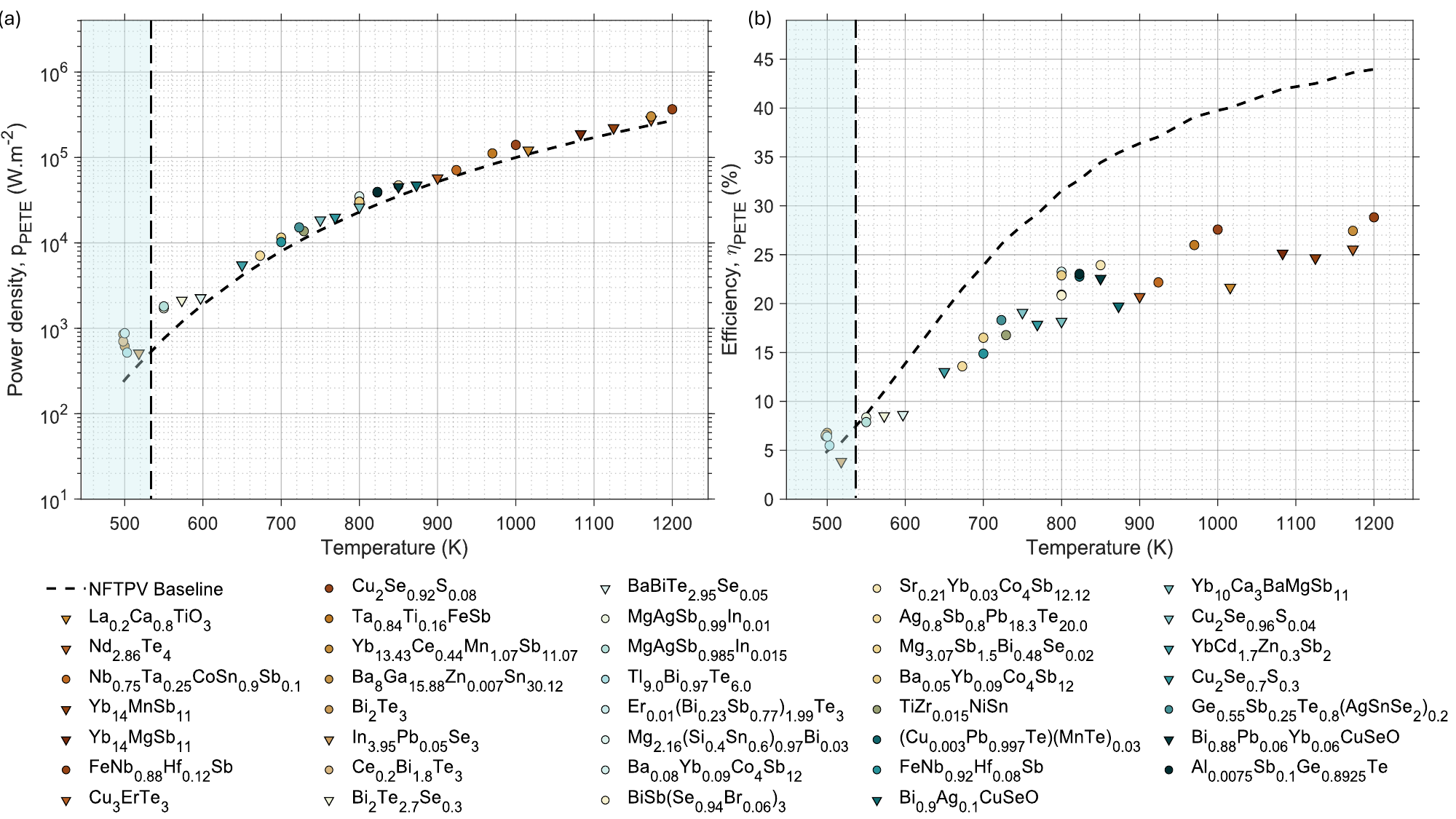}
\caption{Map of different thermoelectric materials used as legs for the PETE device, at different temperatures, compared to standalone NFTPV, at a leg length of 10~nm. (a) Power density as a function of temperature, (b) Efficiency as a function of temperature. For each material, the emitter temperature corresponds to the maximum temperature in the material's dataset. The material properties have been taken from the ESTM dataset \cite{na_public_2022}. The materials with low thermoelectric performance values are distinguished with the triangle symbol.}
\label{fig:figure0.6_scalability_power}
\end{figure}

To understand the extent of the adaptability of our device to temperature, we hybridize the NFTPV cell with 38 different thermoelectric materials \cite{na_public_2022}, chosen under different criteria: (1) high power factor, high overall $ZT$ and (2) low power factor or low $ZT$ to show how material choice affects the performance. The names of the materials as well as the respective PETE performance values can be found in the Supplementary Material (Table~I) \cite{mechlia_supplementary_nodate}. This yields the maps in Fig.~\ref{fig:figure0.6_scalability_power} (a) and (b), showing the resulting performance values and serving as a benchmark and catalog for our technology. These results represent theoretical limits for the specific case of fixed TEG-leg geometries, using the same property symmetry assumption for the $n$- and $p$-type legs as the ideal performance model.
The maps are generated for $FF$ values that produce the mean efficiency value of the PETE device, which is a choice here for demonstration purposes. However, in practice, this leads to a compromise between the goal of the application of the PETE device (power vs. efficiency) and its robustness and durability. Therefore, achieving more than half of the NFTPV efficiency, while achieving a much higher power density and a stable, durable structure is in fact a reasonable performance to aim for.
Another interesting insight that the map presents is that the TEG acts as a "performance buffer" for the PETE performance across all temperature ranges, especially at low temperatures. In this range, we can see how the PETE efficiency is higher than that of the NFTPV, the TEG sometimes outperforms the NFTPV itself, which yields a much higher power density compared to the NFTPV.

\subsection{\label{sec:contact-resistance}Practical Considerations: Contact Resistance}

Contact resistance is a stubborn issue in TEG devices. Many studies report significant power and efficiency losses, reflecting the real-world performance \cite{bjork_universal_2015,li_impacts_2024}.
Furthermore, it has been shown that integrating contact resistance as a lump constant in the model still overestimates the performance and it is suggested that these resistances should be treated separately for the hot and cold junctions, as well as for their position in the device, i.e. internal or external to the TE leg \cite{li_impacts_2024}.

In an effort to approach the real-world behavior, we have included the thermal and electrical contact resistances in the PETE to give realistic estimates and to quantify the loss in performance in comparison with the previous model.
The thermal contact resistances are placed in the same way as suggested in literature \cite{li_impacts_2024} (See Supplementary Material, Fig.~S1a \cite{mechlia_supplementary_nodate}, also see reference \cite{qing_analytical_2018}), where the internal resistances - in between the different layers and the TE leg- are considered equal.
Following the new thermal circuit of the system, the energy conservation equations at the hot and cold junctions are established (See Supplementary Material \cite{mechlia_supplementary_nodate}, also see reference \cite{qing_analytical_2018}), leading to the relationship between the heat flux traversing the thermal resistances and the one entering the leg.

In this study, we are more interested in the order of magnitude of the contact resistances rather than strict values because these depend on the specific contact engineering method and the materials involved. The question that arises in this context: what orders of magnitude for the electrical contact resistance (ECR) and thermal contact resistance (TCR) do we need for a PETE performance comparable to the realistic property model?
To answer this, we started from experimental orders of magnitude of the ECR and the internal and external TCR, which are $10^{-10}\,\Omega\cdot m^2$, $10^{-5}\,K\cdot m^2\cdot W^{-1}$, and $10^{-4}\,K\cdot m^2\cdot W^{-1}$, respectively, reported in literature \cite{li_impacts_2024}. This completely deteriorated the TEG performance: we get the minimal temperature difference set in the optimizer, at 10~K, and a power density orders of magnitude smaller than that of the NFTPV. The efficiency is now completely determined by the NFTPV, which explains its high value for the PETE.
Then, sweeping across the different orders of magnitude of the TCR, then the ECR (Table~\ref{tab:pete_ecr_tcr}), we reach the desired PETE performance at $10^{-15}$ for electrical contact resistance and $10^{-11}$ for the internal thermal contact resistance. The relative effects of the TCR and ECR on the TEG performance are also consistent with the literature \cite{li_impacts_2024}.

To assess the gap between our results and the current state of the art contact resistance values, we made a comparison with the cases using the lowest TCR \cite{cai_interfacial_2026} and ECR \cite{niimi_sub-_2016} achieved, and the same geometric parameters as before.
We found that our target ECR is only one order of magnitude lower than the state of the art, which is $10^{-14}\,\Omega\cdot m^2$ \cite{niimi_sub-_2016}. As for the TCR, we need to further lower the state of the art value, which is $10^{-9}\,K\cdot m^2\cdot W^{-1}$ \cite{cai_interfacial_2026}, by two orders of magnitude. These results imply that the technology could be translated into a manufacturable product in the future.
However, it is important to mention that the state of the art cases were assessed separately by considering the other counterpart to be zero, since they are achieved using different contact engineering methods. Therefore, the challenge is to achieve these orders of magnitude for the same contact engineering methods, and keep refining to reduce the current values, which is quite feasible considering the advancements in contact engineering.

\begin{sidewaystable}[p]
\caption{\label{tab:pete_ecr_tcr}PETE performance under different ECR and TCR orders of magnitude.}
\setlength{\tabcolsep}{2pt}
\renewcommand{\arraystretch}{1.15}
\centering
\resizebox{\textheight}{!}{%
\begin{tabular}{ccccccccccccc}
\hline
\textrm{$r_c$} & \textrm{$r_{th,i},\,r_{tc,i}$} & \textrm{$r_{th,e},\,r_{tc,e}$} & \textrm{$T_{h,n}$} & \textrm{$T_{h,p}$} & \textrm{$T_{c,n}$} & \textrm{$T_{c,p}$} & \textrm{$\Delta T_n$ (K)} & \textrm{$\Delta T_p$ (K)} & \textrm{$P_{\mathrm{TEG}}$ (W$\cdot$m$^{-2}$)} & \textrm{$\eta_{\mathrm{TEG}}$ (\%)} & \textrm{$P_{\mathrm{PETE}}$ (W$\cdot$m$^{-2}$)} & \textrm{$\eta_{\mathrm{PETE}}$ (\%)} \\
\hline
\multicolumn{13}{c}{ECR fixed, internal and external TCR change together} \\
\hline
\multirow{5}{*}{$10^{-10}$} & $10^{-11}$ & $10^{-10}$ & 949.67 & 953.36 & 327.89 & 324.20 & 621.78 & 629.16 & $3.49 \times 10^{1}$ & 0.0078 & $8.74 \times 10^{4}$ & 13.04 \\
 & $10^{-10}$ & $10^{-9}$ & 819.37 & 831.54 & 458.18 & 446.02 & 361.19 & 385.52 & $1.30 \times 10^{1}$ & $4.90 \times 10^{-3}$ & $8.73 \times 10^{4}$ & 17.93 \\
 & $10^{-9}$ & $10^{-8}$ & 673.87 & 678.61 & 603.69 & 598.95 & 70.18 & 79.66 & $5.32 \times 10^{-1}$ & $1.00 \times 10^{-3}$ & $8.73 \times 10^{4}$ & 31.68 \\
 & $10^{-8}$ & $10^{-7}$ & 642.64 & 643.23 & 634.90 & 634.31 & 7.74 & 8.92 & $6.58 \times 10^{-3}$ & $1.00 \times 10^{-4}$ & $8.73 \times 10^{4}$ & 38.13 \\
 & $10^{-7}$ & $10^{-6}$ & 606.52 & 618.71 & 605.52 & 617.71 & 1 & 1 & $9.21 \times 10^{-5}$ & $1.31 \times 10^{-5}$ & $8.73 \times 10^{4}$ & 39.14 \\
\multicolumn{13}{c}{TCR fixed, ECR changes} \\
\hline
$10^{-11}$ & \multirow{5}{*}{$10^{-11}$} & \multirow{5}{*}{$10^{-10}$} & 949.62 & 953.33 & 327.92 & 324.22 & 621.70 & 629.11 & $3.48 \times 10^{2}$ & 0.077 & $8.77 \times 10^{4}$ & 13.04 \\
$10^{-12}$ & & & 949.14 & 953.00 & 328.21 & 324.39 & 620.93 & 628.61 & $3.34 \times 10^{3}$ & 0.73 & $9.07 \times 10^{4}$ & 13.37 \\
$10^{-13}$ & & & 945.78 & 950.78 & 330.22 & 325.54 & 615.56 & 625.24 & $2.40 \times 10^{4}$ & 4.81 & $1.11 \times 10^{5}$ & 15.44 \\
$10^{-14}$ & & & 939.41 & 946.58 & 334.13 & 327.67 & 605.28 & 618.91 & $6.30 \times 10^{4}$ & 10.69 & $1.50 \times 10^{5}$ & 18.52 \\
$10^{-15}$ & & & 937.41 & 945.26 & 335.38 & 328.32 & 602.03 & 616.94 & $7.53 \times 10^{4}$ & 12.15 & $1.63 \times 10^{5}$ & 19.31 \\
\multicolumn{13}{c}{TCR without ECR real-world values} \\
\hline
0 & $10^{-10}$ & $10^{-9}$ & 781.95 & 805.21 & 483.17 & 459.49 & 298.78 & 345.72 & $2.23 \times 10^{4}$ & 6.99 & $1.10 \times 10^{5}$ & 20.28 \\
\multicolumn{13}{c}{ECR without TCR real-world values} \\
\hline
$10^{-14}$ & 0 & 0 & 977.56 & 977.56 & 300 & 300 & 677.56 & 677.56 & $6.63 \times 10^{4}$ & 12.58 & $1.54 \times 10^{5}$ & 19.41 \\
\hline
\end{tabular}
}
\end{sidewaystable}
\clearpage

The scalability and versatility of the PETE make it a very useful device across all industries, such as aerospace, steel and cement industries. It is also compatible with different NFTPV designs, especially with the current advancements in nanofabrication and manufacturing and the maturation of the NFTPV technology.

We add some final notes before concluding this work. First, the $(ZT)_{\mathrm{eng}}$ model assumes that the temperature gradient can be linearized across the TEG leg, which may be true for legs in the millimetric scale, but would need verification in the nanoscale, especially under large temperature differences.
Second, since many applications work at high temperatures, one could argue that using segmented TE legs would deliver a better performance than single-material TE legs. While this is true, the geometrical constraint of the nanoscale gap and the mechanical behavior matching of the segmented materials would make the mass manufacturing of this device less practical. However, this could be further studied, especially with the advancement of nanofabrication techniques.
Third, the thermal and electrical behavior of the TE legs are modeled using the diffuse transport model. At nanoscale leg lengths, this model underestimates performance because phonon and electron transport become decoupled \cite{cimmelli_new_2022, yang_decoupling_2021, smith_macro_2016,xiao_decoupling_2014,heremans_when_2013}. Specifically, since phonons experience strong boundary scattering since their mean free paths span several orders of magnitude larger than the leg dimensions, lattice thermal conductivity is suppressed and therefore conductive heat losses through the legs are reduced. Electrons, on the other hand, are comparatively less affected: for Half-Heusler compounds at temperatures approaching 977~K, electron mean free paths are on the order of 10~nm \cite{zhou_large_2018}, meaning electron transport remains largely preserved at this scale. The net effect is a more favorable ratio of electrical output to thermal loss than what the diffuse model predicts. Quantifying this improvement requires dedicated nanoscale transport modeling and is left for future work.

\section{\label{sec:conclusion}Conclusion}
In this work, we proposed and theoretically validated PETE, a hybrid solid-state architecture in which the passive mechanical spacers of a near-field photonic heat engine are replaced by thermoelectric generators to convert parasitic heat conduction into electricity. This strategy is applicable to any near-field device platform requiring mechanical gap maintenance.
Evaluated on an ITO--InAs near-field thermophotovoltaic platform using TaFeSb- and ZrNiSn-based Half-Heusler thermoelectric materials at a 977~K emitter temperature, PETE roughly doubles the power density of both the NFTPV and the TEG while improving on the standalone efficiency of the TEG by ~44\%. Device performance is tunable through thermoelectric geometry and material selection, and adaptability has been demonstrated across various thermoelectric material pairings spanning a broad temperature range. We also conclude that an order of magnitude of $10^{-15}\,\Omega\cdot m^2$ and $10^{-11}\,K\cdot m^2\cdot W^{-1}$ for the ECR and TCR, respectively, is required for our technology, which is one to two orders of magnitude lower than the state of the art. Future work could incorporate nanoscale quantum transport models into the PETE framework, account for parasitic radiative and convective losses, and extend the analysis to other near-field photonic heat engine platforms.

\section*{Acknowledgments}
The authors acknowledge the Research Computing Data Core at the University of Houston for assistance with the calculations carried out in this work. Y.B.M. thanks Sina Jafari Ghalekohneh for useful discussions about the NFTPV numerical simulations.

\section*{Funding}
This work was supported by the National Science Foundation under Grants CBET-2314210 and CBET-2440814.

\section*{Nomenclature}
\subsection*{Roman Symbols}
\begin{description}
\item[$A_n$, $A_p$] cross-sectional areas of the $n$- and $p$-type legs (m$^2$)
\item[$A_{\mathrm{tot}}$] total radiating surface area of the emitter in the standalone NFTPV device (m$^2$)
\item[$\mathcal{D}_c$] denominator of the power density and heat fluxes ($\Omega$)
\item[$d$] thermoelectric leg length (m)
\item[$f$] thermoelectric leg area ratio
\item[$FF$] fill fraction
\item[$I$] electrical current (A)
\item[$m$] load matching factor
\item[$p_{\mathrm{NFTPV}}$] power density of the standalone NFTPV device (W$\cdot$m$^{-2}$)
\item[$p_{\mathrm{PETE}}$] power density of the PETE device (W$\cdot$m$^{-2}$)
\item[$p_{\mathrm{TEG}}$] power density of the standalone TEG (W$\cdot$m$^{-2}$)
\item[$q_h$] incoming heat flux from the hot side (W$\cdot$m$^{-2}$)
\item[$q_{\mathrm{joule}}$] Joule heating flux (W$\cdot$m$^{-2}$)
\item[$q_{\mathrm{Peltier}}$] Peltier heat flux (W$\cdot$m$^{-2}$)
\item[$q_{\mathrm{rad}}$] radiative heat flux of the NFTPV device (W$\cdot$m$^{-2}$)
\item[$q_{\mathrm{TE}}$] conduction heat flux (W$\cdot$m$^{-2}$)
\item[$R$] internal electrical resistance of the TEG ($\Omega$)
\item[$R_L$] external load resistance ($\Omega$)
\item[$R_n$, $R_p$] electrical resistances of the $n$- and $p$-type legs ($\Omega$)
\item[$R_{n,t,j}$, $R_{p,t,j}$] thermal resistances of the $j$th $n$- and $p$-type legs (K$\cdot$m$\cdot$W$^{-1}$)
\item[$r_c$] electrical contact resistance ($\Omega\cdot$m$^2$)
\item[$r_{n,p}$] integral of the electrical resistivity of the $n$- and $p$-type legs ($\Omega\cdot$m$\cdot$K)
\item[$r_{tc,e}$] external thermal contact resistance on the cold side (K$\cdot$m$^2\cdot$W$^{-1}$)
\item[$r_{tc,i}$] internal thermal contact resistance on the cold side (K$\cdot$m$^2\cdot$W$^{-1}$)
\item[$r_{th,e}$] external thermal contact resistance on the hot side (K$\cdot$m$^2\cdot$W$^{-1}$)
\item[$r_{th,i}$] internal thermal contact resistance on the hot side (K$\cdot$m$^2\cdot$W$^{-1}$)
\item[$S$] Seebeck coefficient ($\mu$V$\cdot$K$^{-1}$)
\item[$T_c$, $T_h$] cold- and hot-junction temperatures (K)
\item[$T_e$] emitter temperature (K)
\item[$T_i$] unknown junction temperature, $i=1,2,3,4$ (K)
\item[$T_r$] receiver (PV cell) temperature (K)
\item[$V_{oc}$] open-circuit voltage (V)
\item[$ZT$] thermoelectric figure of merit
\item[$(ZT)_{\mathrm{eng}}$] figure of merit based on the cumulative-temperature property model
\item[$ZT_{d,\mathrm{eng}}$] device-level figure of merit
\end{description}

\subsection*{Greek Symbols}
\begin{description}
\item[$\Delta T$] temperature difference between the emitter and PV cell, $T_h-T_c$ (K)
\item[$\Delta T_n$] temperature difference between the $n$-type leg junctions, $T_1-T_2$ (K)
\item[$\Delta T_p$] temperature difference between the $p$-type leg junctions, $T_3-T_4$ (K)
\item[$\eta_c$] Carnot efficiency
\item[$\eta_{\mathrm{PETE}}$] PETE device efficiency
\item[$\kappa$] thermal conductivity of the thermoelectric legs (W$\cdot$m$^{-1}\cdot$K$^{-1}$)
\item[$\rho$] electrical resistivity ($\Omega\cdot$m)
\item[$\sigma$] electrical conductivity (S$\cdot$m$^{-1}$)
\item[$\tau$] Thomson coefficient ($\mu$V$\cdot$K$^{-1}$)
\item[$\theta$] integral of the Thomson coefficient ($\mu$V)
\end{description}

\subsection*{Abbreviations}
\begin{description}
\item[ECR] electrical contact resistance
\item[FFTPV] far-field thermophotovoltaics
\item[NEMS] nanoelectromechanical systems
\item[NFTPV] near-field thermophotovoltaics
\item[PETE] photonic and electronic thermoelectric device
\item[PV] photovoltaic
\item[TCR] thermal contact resistance
\item[TE] thermoelectric
\item[TEG] thermoelectric generator
\end{description}

\bibliographystyle{unsrt}
\bibliography{uploads/references}

\end{document}